\documentclass{sf2a-conf2016}
\usepackage{graphicx}
\usepackage{hyperref}
\usepackage[]{natbib}  
\usepackage{epstopdf}

\def\BibTeX{{\rm B\kern-.05em{\sc i\kern-.025em b}\kern-.08em
    T\kern-.1667em\lower.7ex\hbox{E}\kern-.125emX}}
\bibpunct{(}{)}{;}{a}{}{,}  

\begin{document}

\TitreGlobal{SF2A 2016}


\title{Understanding Active Galactic Nuclei using near-infrared high angular resolution polarimetry I : M\lowercase{ont}AGN - STOKES comparison}

\runningtitle{AGN at high angular resolution : MontAGN - STOKES}

\author{L. Grosset}\address{LESIA, Observatoire de Paris, PSL Research University, CNRS, Sorbonne Universit\'es, UPMC Univ. Paris 06, Univ. Paris Diderot, Sorbonne Paris Cit\'e, 5 place Jules Janssen, 92190 Meudon, France}

\author{F. Marin}\address{Observatoire Astronomique de Strasbourg, Universit\'e de Strasbourg, CNRS, UMR 7550, 11 rue de l’Universit\'e, 67000 Strasbourg, France}

\author{D. Gratadour$^1$}
\author{R. Goosmann$^2$}
\author{D. Rouan$^1$}
\author{Y. Cl\'enet$^1$}
\author{D. Pelat}\address{LUTh, Observatoire de Paris, CNRS, Universit\'e Paris Diderot, Sorbonne Paris Cit\'e, 5 place Jules Janssen, 92190 Meudon, France} 
\author{P. Andrea Rojas Lobos$^2$}

\setcounter{page}{237}


\maketitle


\begin{abstract}
In this first research note of a series of two, we present a comparison between two Monte Carlo radiative transfer codes: MontAGN and STOKES. Both were developed in order to better understand the observed polarisation of Active Galactic Nuclei (AGN). Our final aim is to use these radiative transfer codes to simulate the polarisation maps of a prototypical type-2 radio-quiet AGN on a wide range of wavelengths, from the infrared band with MontAGN to the X-ray energies with STOKES. Doing so, we aim to analyse in depth the recent SPHERE/IRDIS polarimetric observations conducted on NGC 1068. In order to validate the codes and obtain preliminary results, we set for both codes a common and simple AGN model, and compared their polaro-imaging results.
\end{abstract}



\section{Introduction}

Polarimetry is a powerful tool as it gives access to more information than spectroscopy or imaging alone, especially about scattering. In particular, indications on the geometry of the distribution of scatterers, the orientation of the magnetic field or the physical conditions can be revealed thanks to two additional parameters : the polarisation degree and the polarisation position angle. Polarimetry can put constraints on the properties of scatterers, like for example spherical grains or oblate grains \citep{Lopez2015} and therefore constrain the magnetic field orientation and optical depth of the medium. The downside is that analysis of polarimetric data is not straightforward. The use of numerical simulations and especially radiative transfer codes is a strong help to understand such data \citep[see for instance][]{Bastien1990,Murakawa2010,Goosmann2011}. It allows us to assess and verify interpretations by producing polarisation spectra/maps for a given structure, which can then be compared to observations.

STOKES and MontAGN are two numerical simulations of radiative transfer both using a Monte Carlo method built to study polarised light travelling through dusty environments (whether stellar or galactic). In both cases, one of the main goal in developing such codes was to investigate the polarisation in discs or tori around the central engine of AGN. While STOKES was designed to work at high energies, from near infrared (NIR) to X rays, MontAGN is optimised for longer wavelength, typically above 1~$\mu$m. Therefore they are covering a large spectral scale with a common band around 0.8 -- 1~$\mu$m. Both approaches are quite different since STOKES is a geometry-based code using defined constant dust (or electrons, atoms, ions ...) three-dimensional 
structures while MontAGN uses a Cartesian 3D grid sampling describing dust densities.

In this first research note, we want to present our first comparison between the two codes. We opted for a similar toy model that we implemented in the two simulation tools in order to produce polarisation maps to be compared one to each other. The second proceedings of this series of two will focus on the results of the code when applied to a toy model of NGC 1068.
  
\section{The radiative transfer Codes}

\subsection{STOKES}

STOKES was initially developed by R. W. Goosmann and C. M. Gaskell in 2007 in order to understand how reprocessing could alter the optical and ultraviolet radiation of radio-quiet AGN \citep{Goosmann2007,Goosmann2007b}. The code was continuously upgraded to include an imaging routine, a more accurate random number generator and fragmentation \citep{Marin2012,Marin2015}, until eventually pushing the simulation tool to the X-ray domain \citep{Goosmann2011,Marin2016}. 
STOKES is a radiative transfer code using Mueller Matrices and Stokes vectors to propagate the polarisation information through emission, absorption and scattering. Photons are launched from a source (or a set of sources) and then propagate in the medium until they are eventually absorbed or they exit the simulation sphere. The optical depth is computed based on the geometry given as an input. At each encounter with a scatterer, the photon's absorption is randomly determined from the corresponding albedo; if it is absorbed, another photon is launched from the central source. In a scattering case, the new direction of propagation is determined using phase functions of the scatterer and the Stokes parameters are modified according to the deviation. For a detailed description of the code, see papers of the series \citep{Goosmann2007,Marin2012,Marin2015}.

\subsection{MontAGN}

Following the observation of NGC 1068 in polarimetric mode at high angular resolution conducted by \cite{Gratadour2015}, MontAGN (acronym for ``Monte Carlo for Active Galactic Nuclei'') was developed to study whether our assumptions on the torus geometry were able to reproduce the observed polarisation pattern through simulations in the NIR. MontAGN has many common points with STOKES. Since the two codes were not designed for the same purpose, the main differences originate from the effects that need to be included in the two wavelength domains, which differ between the infrared and the shorter wavelengths. STOKES includes Thomson scattering, not available in MontAGN, while MontAGN takes into account the re-emission by dust as well as temperature equilibrium adjustment at each absorption to keep the cells temperature up to date, not present in STOKES.

In MontAGN photons are launched in the form of frequency-independent photon packets. If absorption is enabled, when a photon packet is absorbed, it is immediately re-emitted at another wavelength, depending on the dust temperature in the cell. The cell temperature is changed to take into account this incoming energy. The re-emission depends on the difference between the new temperature of  the cell and the old one to correct the previous photon emissions of the cell at the former temperature (following \citealt{Bjorkman2001}). If re-emission is disabled, all photon packets are just scattered, but we apply the dust albedo as a factor to the energy of the packet to solely keep the non-absorbed fraction of photons (see \citealt{Murakawa2010}). This disabling allows us to get much more statistics at the end of the simulation as every photon is taken into account. But it also requires to have a lot of photons in each pixel at the end as we may obtain in one pixel only photons with weak probability of existence, a situation that is not representative of the actual pixel polarisation.




\section{Simulation}

\begin{figure}[ht!]
 \centering
 \includegraphics[width=0.30\textwidth,clip]{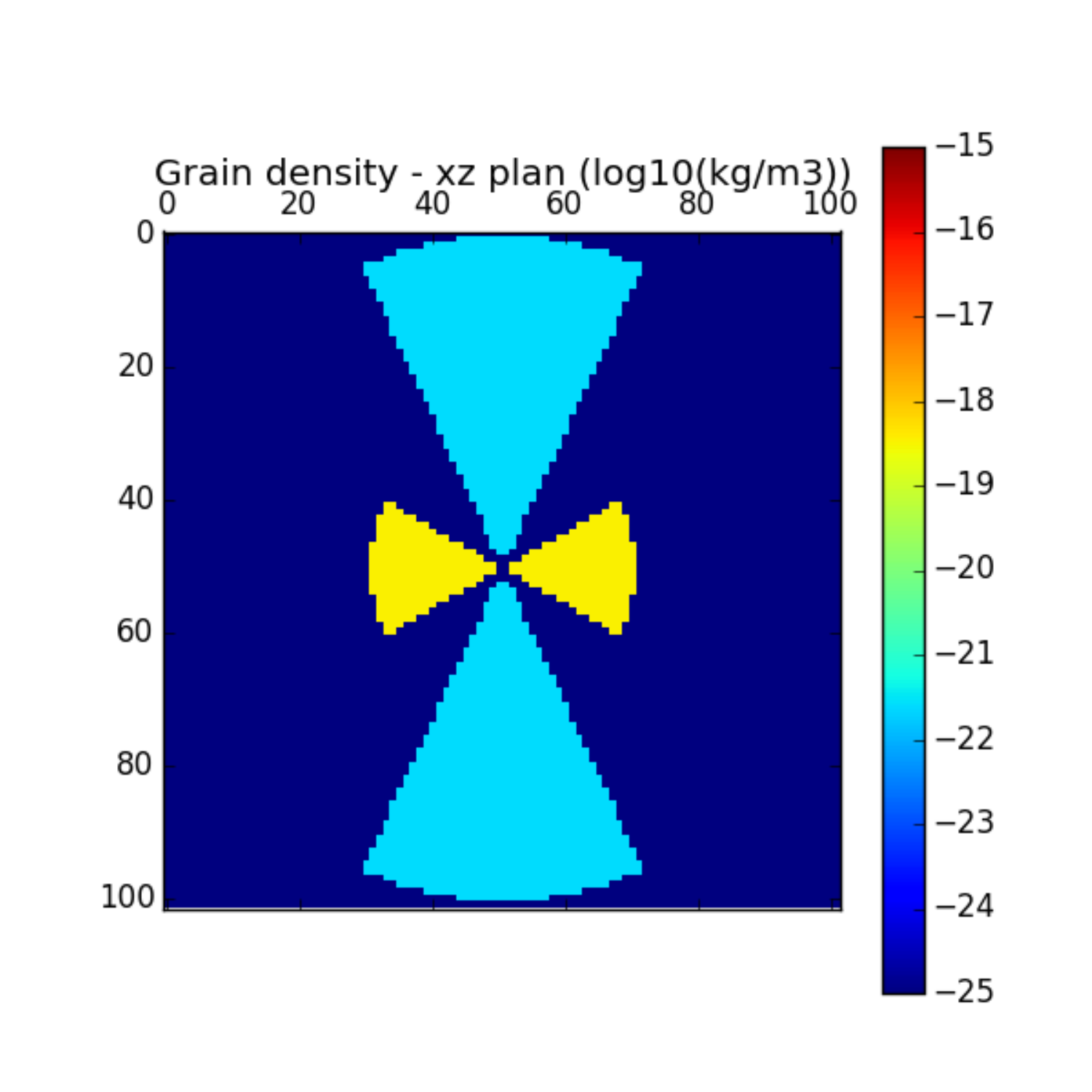}%
 \includegraphics[width=0.30\textwidth,clip]{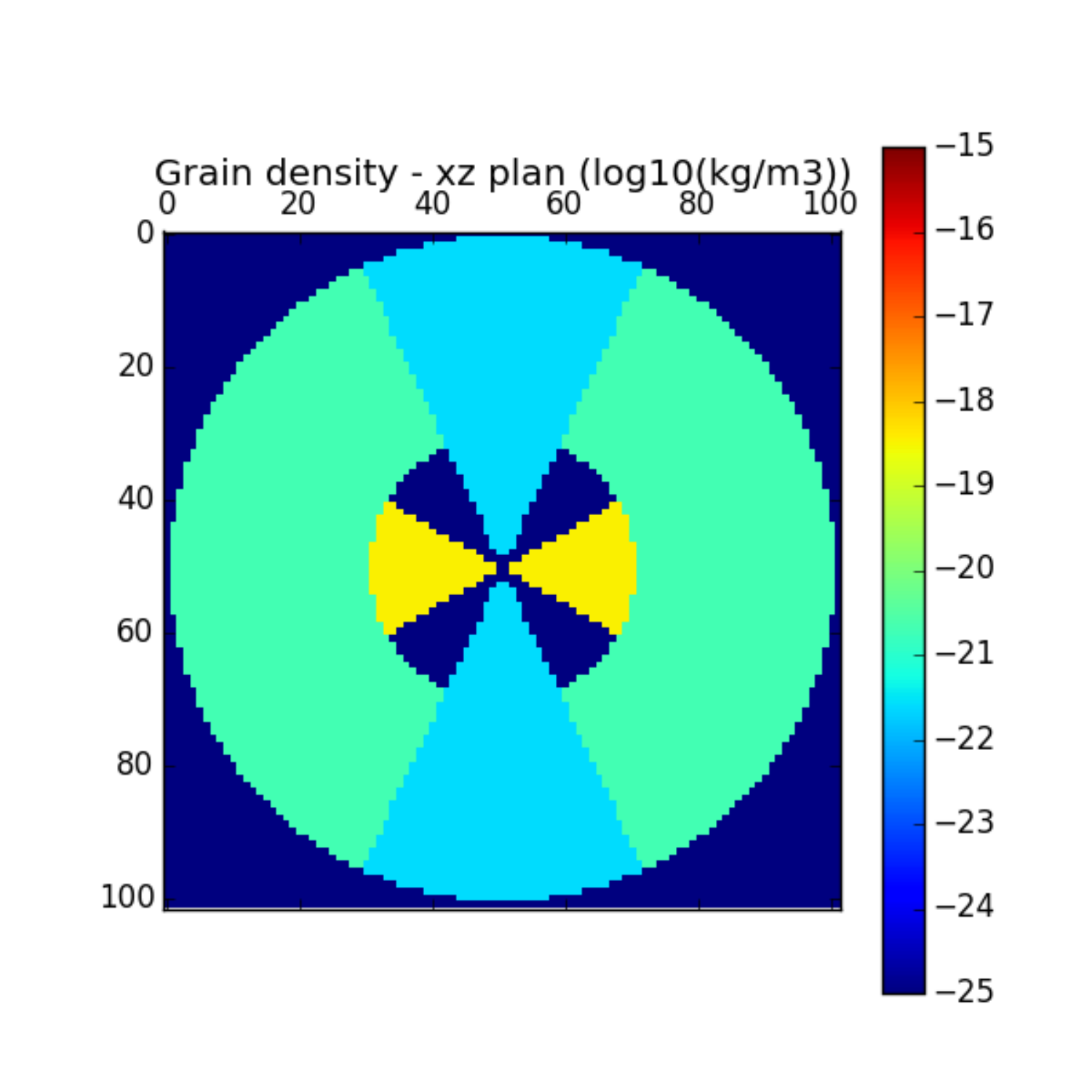} 
  \caption{Grain density (in kg/m$^3$) set for both models. Note that in STOKES the polar outflow is constituted of electrons, at a density allowing us to have the same optical depth {\bf Left:} first model : "model I" {\bf Right:} second model with the dust shell : "model II".}
  \label{model}
\end{figure}

We set up a model of dust distribution compatible with the two codes. At the centre of the model, a central, isotropic, point-like source is emitting unpolarised photons at a fixed wavelength (0.8, 0.9 and 1~$\mu$m, only images at 0.9~$\mu$m are shown in this publication). Around the central engine, is a flared dusty disk with radius ranging from 0.05~pc to 10~pc. It is filled with silicate grains and has an optical depth in the V-band of about 50 along the equatorial plane (see Fig.~\ref{model}). Along the polar direction, a bi-conical, ionised wind with a 25$^\circ$ half-opening angle with respect to the polar axis flows from the central source up to 25 pc. The wind is filled with electrons in STOKES and silicate grains at much lower density in MontAGN\footnote{This difference in composition should not affect the polarisation results as the two corresponding phase functions are close, but the flux and polarised flux will be attenuated in the case of dust because of absorption. Thomson scattering should therefore be included in MontAGN to be more realistic.}. The conical winds are  optically thin ($\tau _{V}$ = 0.1). We added to these structures a cocoon of silicate grains surrounding the torus, from 10 pc to 25 pc, outside the wind region to account for a simplified interstellar medium in another model. See our second proceedings of this series (Marin, Grosset et al., hereafter Paper~II) for more information about the models. Re-emission was disabled for MontAGN in these simulations.

\begin{figure}[ht!]
 \centering
 \includegraphics[width=0.30\textwidth,clip]{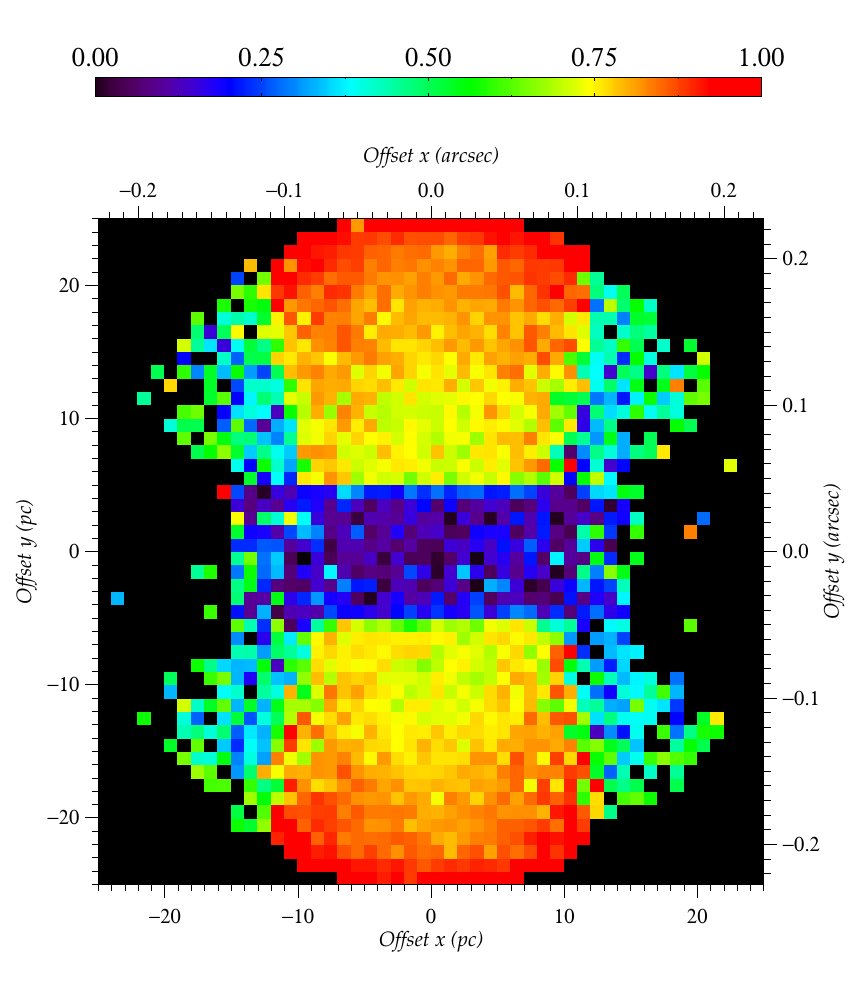}
 \includegraphics[width=0.39\textwidth,clip]{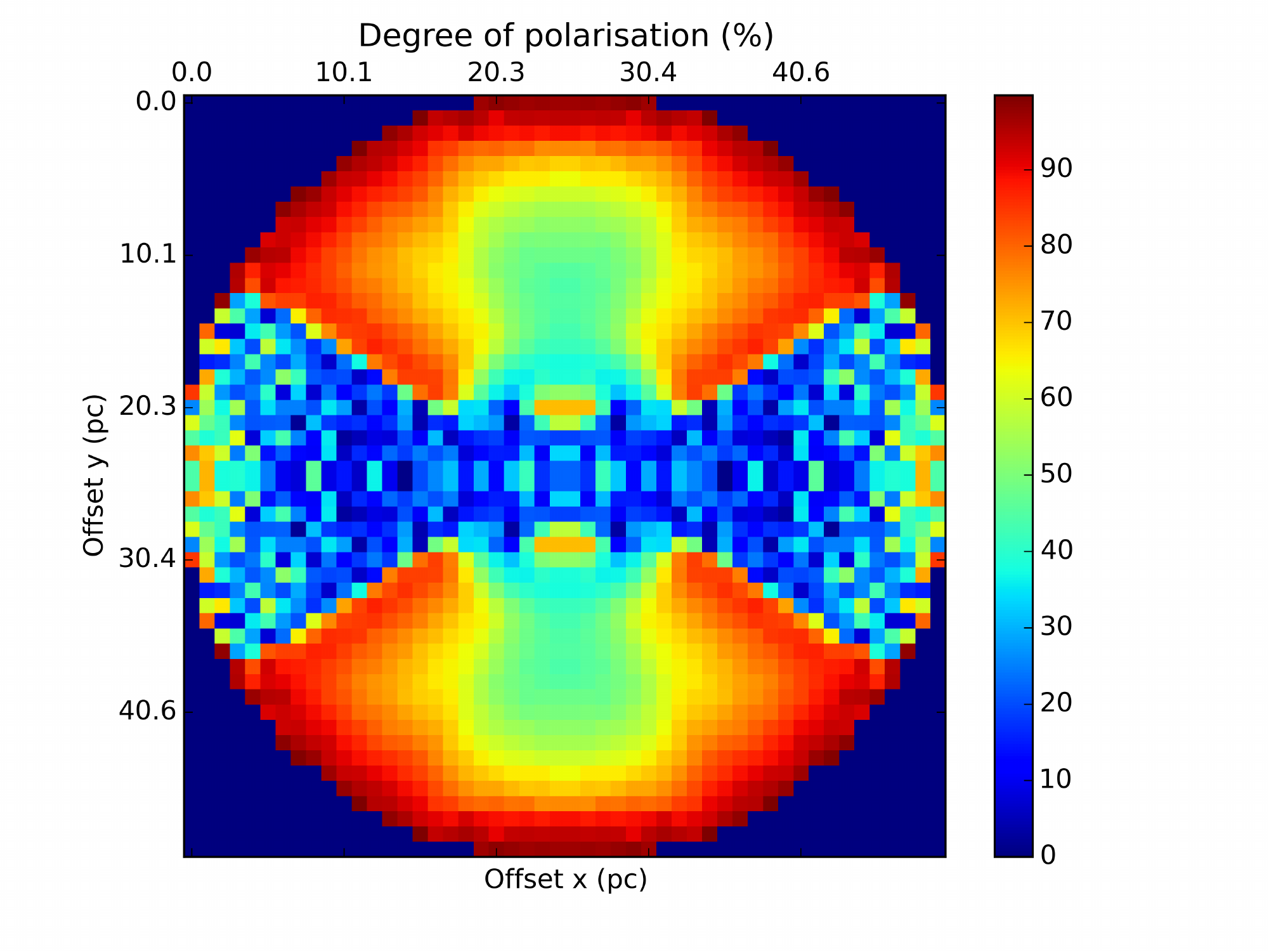}
  \caption{Polarisation degree for model II at 0.9 $\mu$m {\bf Left:} with STOKES {\bf Right:} with MontAGN (in $\%$)}
  \label{modI_P}
\end{figure}

\begin{figure}[ht!]
 \centering
 \includegraphics[width=0.30\textwidth,clip]{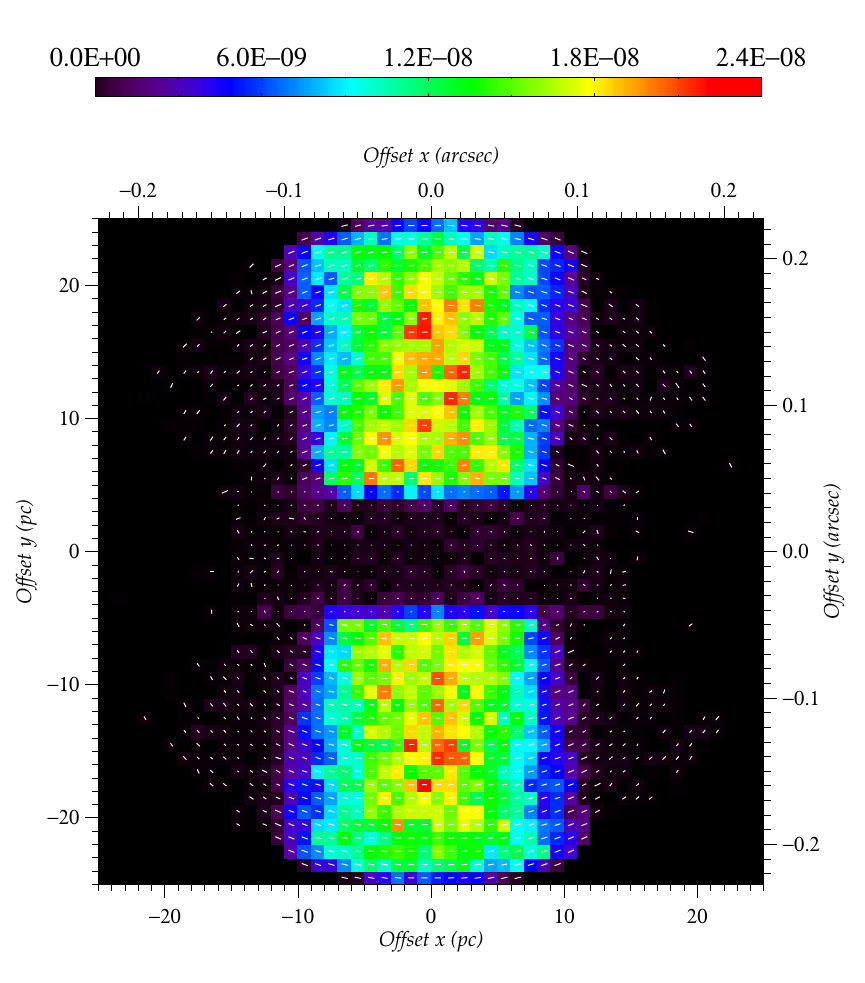}
 \includegraphics[width=0.39\textwidth,clip]{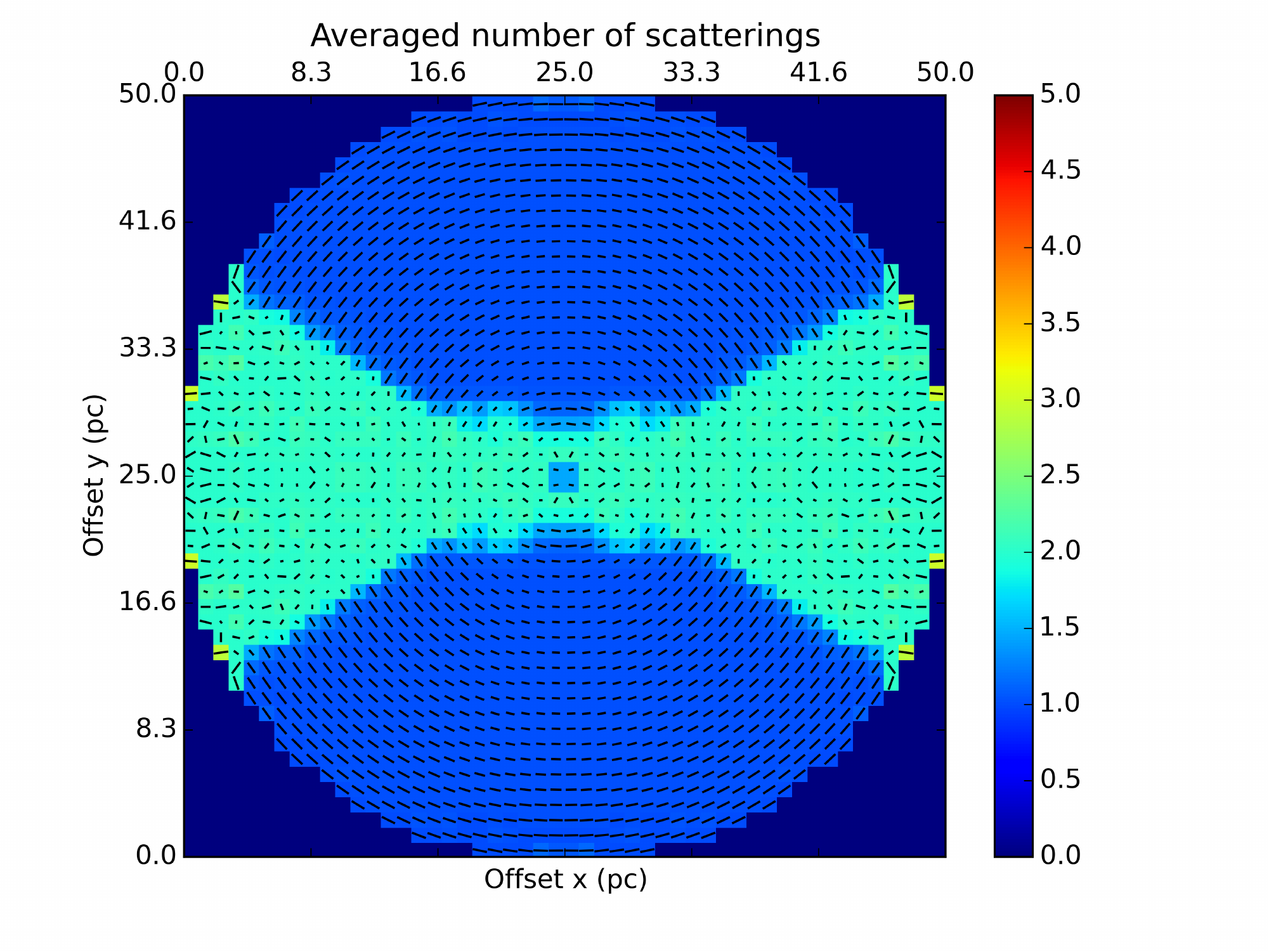}
  \caption{At 0.9~$\mu$m {\bf Left:} Polarised flux (in arbitrary units) with the polarisation position angle superimposed for model II with STOKES. {\bf Right:} Averaged number of scatterings a photon undergo before exiting the medium with polarisation vectors superimposed for model II with MontAGN. }
  \label{modI_vect}
\end{figure}



With more than 5$\times$10$^6$ photons sampled, we obtain for both models an overall good agreement between the two codes. In the polar outflow region, the similarities are high between the two codes, revealing high polarisation degrees (close to 100\%) despite the differences in composition (see Fig.~\ref{modI_P}). This is expected from single scattered light at an angle close to 90$^\circ$ (see \citealt{Bastien1990}), which is confirmed from the maps of averaged number of scatterings (see Fig.~\ref{modI_vect}, right). However in the central region, where the torus is blocking the observer's line-of-sight, the results between MontAGN and STOKES slightly differ (see the equatorial detection of polarisation at large distances from the centre in Fig.~\ref{modI_P}, right). We interpret this polarisation as arising from the differences in the absorption method between the two codes. Because in MontAGN all photons exit the simulation box, we always get some signal even if it may not be representative of photons reaching this peculiar pixel. If inside a pixel only photons with low probability, i.e. with the energy of their photon packets being low after multiple scatterings, are collected, the polarisation parameters reconstructed from these photons will not be reliable. This is why we need to 
collect an important number of photons per pixel. 

Otherwise, the polarisation structure revealed by polaro-imaging is very similar between the two codes and lead to distinctive geometrical highlights that will be discussed in the second research note of this series (Paper II).

\section{Concluding remarks}

We compared the MontAGN and STOKES codes between 0.8 and 1~$\mu$m for similar distribution of matter and found that many of the polarimetric features expected from one code are reproduced by the second. The only difference so far resides on the detection of polarisation at large distances from the centre of the model, where we need a higher sampling in order for MontAGN to match STOKES results. The next step will be to improve the models, and develop the MontAGN code by including more effects like electron scattering or non spherical grains (ortho- and para-graphite, namely). However we already get a fairly good agreement between the codes which give us confidence to pursue our exploration of the near-infrared signal of AGN together with MontAGN and STOKES. Note that the comparison allowed to detect flaws and bugs, a positive outcome. We intend to explore our first results in Paper~II and push the codes towards more complex geometries. Once a complete agreement will be found in the overlapping band (0.8 -- 1~$\mu$m), we will run a large simulation ranging from the far-infrared to the hard X-rays for a number of selected radio-quiet AGN. Our targets include the seminal type-2 NGC~1068, as well as a couple of other nearby AGN with published polarimetric data. Forthcoming new infrared polarimetric observations using SPHERE will complement our database and be modelled with MontAGN and STOKES.



\begin{acknowledgements}
The authors would like to acknowledge financial support from the Programme National Hautes Energies (PNHE).
\end{acknowledgements}

\bibliographystyle{aa}  
\bibliography{grosset} 

\end{document}